\documentclass{article}
\usepackage{ltwol2e}

\input{psfig.sty}

\def\Journal#1#2#3#4{{#1} {\bf #2}, #3 (#4)}

\def\NPB{{\em Nucl. Phys.} B}
\def\PLB{{\em Phys. Lett.}  B}
\def\PRL{\em Phys. Rev. Lett.}
\def\PRD{{\em Phys. Rev.} D}
\def\PREP{{\em Phys. Rep.} }

\def\IJMP{{\em Int. J. Mod. Phys.} A}

\def\delmchi{\Delta m_{\widetilde \chi}}

\def\bit{\begin{itemize}}
\def\eit{\end{itemize}}
\def\beq{\begin{equation}}
\def\eeq{\end{equation}}
\def\bea{\begin{eqnarray}}
\def\eea{\end{eqnarray}}

\def\stau{\wtil\tau}
\def\mx{M_X}

\def\mz{m_Z}

\def\h{h}
\def\mh{m_{\h}}
\def\mhmax{\mh^{\rm max}}
\def\mhmin{\mh^{\rm min}}

\def\lam{\lambda}

\def\gtino{\wt G}
\def\mgtino{m_{\gtino}}
\def\mplanck{M_{\rm P}}
\def\mpl{\mplanck}

\def\rpm{R^{\pm}}

\def\rzero{R^0}

\def\glsp{$\wtil g$-LSP}
\def\lsim{\mathrel{\raise.3ex\hbox{$<$\kern-.75em\lower1ex\hbox{$\sim$}}}}
\def\gsim{\mathrel{\raise.3ex\hbox{$>$\kern-.75em\lower1ex\hbox{$\sim$}}}}
\def\ifmath#1{\relax\ifmmode #1\else $#1$\fi}

\def\ejet{E_{\rm jet}}
\def\thetamuid{\theta(\mu\mbox{id})}

\def\vev#1{\langle #1 \rangle}
\def\lam{\lambda}

\def\mplanck{M_{\rm Planck}}

\def\sq{\wt q}

\def\msq{m_{\sq}}

\def\slepl{\wt \ell_L}
\def\mslepl{m_{\slepl}}
\def\slepr{\wt \ell_R}
\def\mslepr{m_{\slepr}}

\def\staur{\wt \tau_R}

\def\delgs{\delta_{GS}}

\def\slash#1{#1\hskip-4pt/\hskip0pt}
\def\ptmiss{\slash p_T}

\def\susyslash{\susy\hskip-18pt/\hskip13pt}

\def\mhalf{m_{1/2}}
\def\gl{\wt g}
\def\mgl{m_{\gl}}

\def\etc{{\it etc.}}

\def\sign{{\rm sign}}

\def\h{h}

\def\mh{m_{\h}}

\def\etc{{\em etc.}}

\def\eg{{\it e.g.}}
\def\etal{{\it et al.}}
\def\mhalf{m_{1/2}}
\def\dmm{\Delta^{--}}
\def\dm{\Delta^{-}}
\def\mdmm{m_{\dmm}}
\def\hdmm{h^{\dmm}}
\def\dpp{\Delta^{++}}
\def\delp{\Delta^{+}}

\def\hzero{\Delta^0}

\def\gamdmm{\Gamma_{\dmm}}

\def\stopone{\wt t_1}
\def\stoptwo{\wt t_2}

\def\mstoptwo{m_{\stoptwo}}

\def\sq{\wt q}

\def\msq{m_{\sq}}

\def\slepl{\wt \ell_L}
\def\mslepl{m_{\slepl}}
\def\slepr{\wt \ell_R}
\def\mslepr{m_{\slepr}}

\def\susy{{\rm SUSY}}

\def\etc{{\it etc.}}

\def\sign{{\rm sign}}

\def\etc{{\em etc.}}

\def\eg{{\it e.g.}}
\def\etal{{\it et al.}}
\def\mhalf{m_{1/2}}
\def\gl{\wt g}
\def\mgl{m_{\gl}}

\def\stopone{\wt t_1}

\def\sq{\wt q}

\def\msq{m_{\sq}}

\def\slepl{\wt \ell_L}
\def\mslepl{m_{\slepl}}
\def\slepr{\wt \ell_R}
\def\mslepr{m_{\slepr}}

\def\hsm{h_{\rm SM}}
\def\mhsm{m_{\hsm}}
\def\hl{h^0}

\def\ha{A^0}

\def\mhl{m_{\hl}}

\def\mha{m_{\ha}}

\def\tanb{\tan\beta}

\def\mt{m_t}

\def\mz{m_Z}

\def\mgut{M_U}
\def\mx{M_X}

\def\wm{W^-}

\def\cnone{\wt\chi^0_1}
\def\cntwo{\wt\chi^0_2}

\def\snu{\wt\nu}

\def\msnu{m_{\snu}}
\def\mcnone{m_{\cnone}}
\def\mcntwo{m_{\cntwo}}

\def\h{h}
\def\mh{m_{\h}}
\def\wt{\widetilde}
\def\cpone{\wt \chi^+_1}
\def\cmone{\wt \chi^-_1}
\def\cpmone{\wt \chi^{\pm}_1}
\def\mcpone{m_{\cpone}}
\def\mcpmone{m_{\cpmone}}

\def\emem{e^-e^-}
\def\dmm{\Delta^{--}}
\def\mdmm{m_{\dmm}}

\def\dpp{\Delta^{++}}
\def\delp{\Delta^{+}}

\def\hzero{\Delta^0}

\def\gamdmm{\Gamma_{\dmm}}

\def\wtil{\widetilde}

\def\lam{\lambda}
\def\br{BF}

\def\gam{\gamma}

\def\etal{{\it et al.}}
\def\etc{{\it etc.}}

\def\anti{\overline}
\def\epem{e^+e^-}

\def\mupmum{\mu^+\mu^-}

\def\rts{\sqrt s}
\def\ie{{\it i.e.}}
\def\eg{{\it e.g.}}

\def\anti{\overline}

\def\wm{W^-}

\def\mz{m_Z}
\def\h{h}
\def\mh{m_{\h}}

\def\hsm{h_{SM}}
\def\mhsm{m_{\hsm}}

\def\tanb{\tan\beta}
\def\hl{h^0}
\def\mhl{m_{\hl}}

\def\ha{A^0}
\def\mha{m_{\ha}}

\def\fbi{~{\rm fb}^{-1}}
\def\fb{~{\rm fb}}
\def\pbi{~{\rm pb}^{-1}}

\def\gev{~{\rm GeV}}
\def\tev{~{\rm TeV}}

\def\mt{m_t}

\def\dmm{\Delta^{--}}
\def\mdmm{m_{\dmm}}
\def\hdmm{h^{\dmm}}
\def\dpp{\Delta^{++}}

\def\hzero{\Delta^0}

\def\gamdmm{\Gamma_{\dmm}}

\def\overlay#1#2{\ifmmode \setbox 0=\hbox {$#1$}\setbox 1=\hbox to\wd 0{\hss
$#2$\hss }\else \setbox 0=\hbox {#1}\setbox 1=\hbox to\wd 0{\hss #2\hss }\fi
#1\hskip -\wd 0\box 1}
\def\case#1/#2{{\textstyle{#1\over#2}}}
\def\9{\phantom 0}      
\renewcommand\linebreak{\unskip\break} 
\newcommand{\alt}{\mathrel{\raisebox{-.6ex}{$\stackrel{\textstyle<}{\sim}$}}}
\newcommand{\agt}{\mathrel{\raisebox{-.6ex}{$\stackrel{\textstyle>}{\sim}$}}}
\def\lsim{\alt}
\def\gsim{\agt}
\begin{document}
\font\fortssbx=cmssbx10 scaled \magstep2
\begin{table*}
\hbox to \hsize{
$\vcenter{
\hbox{\fortssbx University of California - Davis}
}$
\hfill
$\vcenter{
\hbox{\bf UCD-98-17} 
\hbox{\bf hep-ph/9810394}
\hbox{October, 1998}
}$
}
\vskip 1in
\centerline{\bf SELECTED LOW-ENERGY SUPERSYMMETRY PHENOMENOLOGY
TOPICS$^\dagger$}
\vskip .25in
\centerline{J. F. Gunion}
\vskip .25in
\centerline{Department of Physics, University of California at Davis,
Davis, CA 95616}
\vskip .75in
\noindent Abstract: I review selected topics in supersymmetry,
including:
effects of non-universality, high $\tanb$ and phases on SUSY signals;
a heavy gluino as the LSP;
gauge-mediated SUSY signals involving delayed decays;
R-parity violation and the very worst case for SUSY discovery;
some topics regarding Higgs bosons in supersymmetry; and
doubly-charged Higgs and higgsinos in supersymmetric left-right symmetric
models. I emphasize scenarios in which detection
of supersymmetric particles and/or the SUSY Higgs bosons might 
require special experimental and/or analysis techniques.
\vskip 3in

$^\dagger$~{Presented at
          the XXIX International Conference on High Energy Physics,
          Vancouver, BC, July 23 -- July 29, 1998.}

\end{table*}
\clearpage

\title{SELECTED LOW-ENERGY SUPERSYMMETRY PHENOMENOLOGY TOPICS}

\author{J. F. Gunion}

\address{Department of Physics, University of California at Davis,
Davis, CA 95616, USA\\E-mail: jfgucd@ucdhep.ucdavis.edu}


\twocolumn[\maketitle\abstracts{I review selected topics in supersymmetry,
including:
effects of non-universality, high $\tanb$ and phases on SUSY signals;
a heavy gluino as the LSP;
gauge-mediated SUSY signals involving delayed decays;
R-parity violation and the very worst case for SUSY discovery;
some topics regarding Higgs bosons in supersymmetry; and
doubly-charged Higgs and higgsinos in supersymmetric left-right symmetric
models. I emphasize scenarios in which detection
of supersymmetric particles and/or the SUSY Higgs bosons might 
require special experimental and/or analysis techniques.}]

\section{\boldmath Non-universality, $\tanb\gg1$ and phases}

Many deviations from universal boundary conditions
at the unification or string scale are now being actively considered.
Neither the gaugino masses nor the scalar masses are required
to be universal. One well-motivated model with non-universal
gaugino masses is the O-II orbifold model,~\cite{cdg}
in which supersymmetry breaking ($\susyslash$)
is dominated by the overall size modulus (as opposed to the dilaton).
It is the only string model where the limit of pure modulus $\susyslash$ is
possible without charge and/or color breaking. One finds.
\beq
M_3:M_2:M_1\stackrel{\scriptstyle O-II}{\sim}-(3+\delgs):(1-\delgs):
({33\over 5}-\delgs)\,,
\label{oiibc}
\eeq
The phenomenology of this model changes dramatically as a function
of the Green-Schwarz parameter, $\delgs$; indeed,
a heavy gluino is the LSP when $\delgs\sim -3$ (a preferred
range for the model). 
Another class of models with non-universal gaugino masses are those
where $\susyslash$ arises due
to $F$-term breaking with $F\neq$ SU(5) singlet.~\cite{snowtheory2}
Possible representations for $F$ include:
\begin{equation}
F\in ({\bf 24}{\bf \times} 
{\bf 24})_{\rm symmetric}={\bf 1}\oplus {\bf 24} \oplus {\bf 75}
 \oplus {\bf 200}\,,\nonumber
\label{irrreps}
\end{equation}
leading to
$\langle F \rangle_{ab}=c_a\delta_{ab}$, with $c_a$ depending
on the representation. Results for the gauginos masses at the
grand-unification scale $\mgut$ and at $\mz$ are given
in Table~\ref{masses}.

\begin{table}[h]
\begin{center}
\begin{tabular}{|c|ccc|ccc|}
\hline
\ & \multicolumn{3}{c|} {$\mgut$} & \multicolumn{3}{c|}{$\mz$} \cr
$F$ 
& $M_3$ & $M_2$ & $M_1$ 
& $M_3$ & $M_2$ & $M_1$ \cr
\hline 
${\bf 1}$   & $1$ &$\;\; 1$  &$\;\;1$   & $\sim \;6$ & $\sim \;\;2$ & 
$\sim \;\;1$ \cr
${\bf 24}$  & $2$ &$-3$      & $-1$  & $\sim 12$ & $\sim -6$ & 
$\sim -1$ \cr
${\bf 75}$  & $1$ & $\;\;3$  &$-5$      & $\sim \;6$ & $\sim \;\;6$ & 
$\sim -5$ \cr
${\bf 200}$ & $1$ & $\;\; 2$ & $\;10$   & $\sim \;6$ & $\sim \;\;4$ & 
$\sim \;10$ \cr
\hline
 $\stackrel{\textstyle O-II}{\delgs=-4}$ & $1$ & $\;\;5$ & ${53\over 5}$ & 
$\sim 6$ & $\sim 10$ & $\sim {53\over5}$ \cr
\hline
\end{tabular}
\caption{$M_a$ at $\mgut$ and $\mz$
for the four $F$ irreducible representations
and in the O-II model with $\delgs\sim -4$.}
\label{masses}
\end{center}
\end{table}

Both $F\in {\bf 200}$ and the O-II model allow for the
possibility that $\mcpmone\simeq\mcnone$, since
$\mcnone\sim {\rm min}(M_1,M_2)$, $\mcpmone\sim M_2$ and $M_2<M_1$.
In this situation, there are two possibilities.  (1) The degeneracy
is so extreme ($\delmchi\equiv \mcpmone-\mcnone\lsim 0.1\gev$) that 
the $\cpmone$ is long-lived. 
In this case, one searches for heavily-ionizing charged tracks.
(2) The degeneracy is still small, but 
large enough that the $\cpmone$ is not pseudo-stable: $0.3\leq \delmchi\leq
3\gev$.
One must search for $\cpone\cmone$ production at an $\epem$ collider
using a photon tag: $\epem\to
\gam\cpone\cmone$. In case (1) [(2)], a DELPHI 
analysis~\cite{delphideg} yields $\mcpmone\geq 84\gev$ [$\geq 54\gev$, provided
$\msnu$ is large]. 
In general (but not preferred in the GUT context),
there is also a third possibility: $M_2,M_1\gg |\mu|$.
In this case, the $\cpone$ and $\cnone$ are again nearly degenerate,
but the $\gam\cpmone\cmone$ cross section is smaller and no limits
(beyond the LEP $\mcpmone\geq 45\gev$ limit)
are possible unless the $\cpmone$ is pseudo-stable.~\cite{delphideg}

Scalar mass non-universality can emerge from many sources;
a particularly popular source is 
$D$-term contributions to scalar mass,
especially from an anomalous U(1). 
A typical model is one~\cite{gmur} 
which employs U(1)$_Y$. The result
is a Fayet-Illiopoulos $D$-term contribution to the scalar masses at $\mgut$:
$\wtil m_i^2=m_0^2+Y_iD_Y$, where $m_0$ is
the usual mSUGRA universal mass. (The other
mSUGRA parameters are denoted $\mhalf$, $A_0$, $\tanb$, $\sign(\mu)$.)
As $D_Y$ is turned on, the scalar masses are altered and the
value of $|\mu|$ required for RGE electroweak symmetry breaking 
(in which $m_{H_2^0}^2$, the scalar
mass-squared associated with the Higgs boson that couples to the top quark,
becomes negative at low energy scales)
to give the correct value of $\mz$ changes.
The `normal' mSUGRA relation between
gaugino masses, scalar masses and $|\mu|$ is altered so that the 
LSP need not be the $\cnone$.
As $D_Y$ is changed, it becomes possible for the LSP to be:
the $\staur$ ($|\mu|>|\mu|_{mSUGRA}$);
a higgsino ($|\mu|<|\mu|_{mSUGRA}$); or a
sneutrino (in a small band with $|\mu|<|\mu|_{mSUGRA}$).
Cosmology suggests these latter are disfavored, but reheating can obviate
such constraints and even a stable LSP=$\staur$ would then be allowable.

Clearly, such scalar non-universality leads to 
drastic changes in collider phenomenology.  In particular,
if the $\staur$ is the LSP one should look for a stable $\staur$,
whereas if a higgsino is the LSP then $\mcpmone\simeq\mcnone\simeq\mcntwo$ 
and LEP2 constraints will be weakened (see above).
Further, in collider events there will be much less 
missing transverse momentum ($\ptmiss$) than for mSUGRA
boundary conditions.

Let us next mention the phenomenological implications of high $\tanb$
for superparticle discovery.
RGE equations cause $\wtil \tau$ to decline in mass relative to 
$\wtil e,\wtil\mu$ (but the $\cnone$ is still the LSP).
This leads to dominance of
$\tau$'s in cascade decays and in the `tri-lepton' signal.
Tevatron signals for SUSY become more difficult; it definitely takes
TeV33 to probe SUSY if gluino and squarks are $\gsim 1\tev$ with
corresponding mass scales for other sparticles.~\cite{bct}

Normally, the possible phases for the soft-SUSY-breaking parameters
have been neglected in studying SUSY collider phenomenology.
For example, in mSUGRA, $A_0$ and $\mu$ can have phases. 
More generally, there are 79
masses and real mixing angles and 45 CP-violating phases in the MSSM.
These phases appear in mass matrices as well as couplings.
EDM and CP-violation constraints
do not require that these phases be small; cancellations among different
contributions to CP-violating observables are possible.~\cite{nathibrahim}
Extraction of all SUSY parameters from experiment
becomes considerably more complex in general,~\cite{kb}
even at an $\epem$ collider.

\section{\boldmath A heavy gluino as the LSP}

There are several attractive models in which the gluino is heavy
and yet is the LSP.  These models include:
the O-II model discussed earlier~\cite{cdg} when 
$\delgs\sim -3$ (the preferred range);
and the GMSB model of Raby.~\cite{rabytobe} 

A detailed study of the phenomenology of a \glsp\ has appeared.~\cite{bcg}
First, one must consider constraints coming from the relic density of
$\rzero=\gl g$ (almost certainly the lightest) bound states.
Taking into account annihilations that continue after freezeout,
and allowing for non-perturbative contributions
to the annihilation cross section, it is found~\cite{bcg} that
the relic density can be small enough, even at very large $\mgl$
and even without including late stage inflation (as might
be needed for the Polonyi problem), to avoid all constraints
from stable isotope searches, underground detectors, \etc\
Certainly, the $\rzero$'s are very unlikely to
be the primary halo constituent.

Next, one must consider how the \glsp\ manifests itself
in a detector and in relevant experimental analyses.
This is sensitively dependent upon several ingredients.
First, there is the question of how the $\gl$ hadronizes.
In general, it can pick up quarks and/or a gluon to form either charged
$\rpm$ (\eg\ $\gl u\anti d$) or neutral $\rzero$ (\eg\ $\gl g, \gl u\anti
u,\ldots$) bound states with probabilities $P$ and $1-P$, respectively.
($\rpm$ states that are not pseudo-stable between hadronic collisions
are not counted in $P$.) These probabilities are assumed
to apply to a heavy $\gl$ both as it exits from the 
initial hard interaction and also after each hadronic collision. (The
picture is that the light quarks and gluons are stripped away
in each hadronic collision and that the heavy $\gl$ is then
free to form the $\rpm$ and $\rzero$ bound states in the same
manner as after initial production.) In any reasonable quark-counting model
$P<1/2$, in which case the $\gl$ spends most of its time
as an $\rzero$ as it passes through the detector.
The second critical ingredient is the
$\vev{\Delta E}$  deposited in a hadronic collision;
several models that bracket the known result for a pion are employed.
Since, a heavy $\gl$ is typically not produced with an ultra-relativistic
velocity, it does not deposit very much energy even in its first
few hadronic collisions; indeed,
it can often penetrate the detector unless it is in an $\rpm$ state
a large fraction of the time ($P>1/2$) and is slowed down by ionization energy
deposits. 
Third, the net hadronic energy deposit 
depends on $\lam_T$, the path length
in iron given by the $\gl$ {\it total} cross section.
One popular model~\cite{gs} suggests $\lam_T\sim 2\lam_T(\pi)$.
Fourth, the 
effective Fe thickness of instrumented and uninstrumented portions of 
the relevant detectors (OPAL and CDF) must be known.
Fifth, one must account for
how a calorimeter treats ionization energy deposits 
as compared to hadronic collision energy deposits; the latter are
measured correctly when the calorimeter is
calibrated for a light hadron, but the former are over-estimated
by a factor of roughly 1.6 in an iron calorimeter (as employed
by OPAL and CDF). Thus,
when a calorimeter is calibrated to give correct $\pi$ energy,
calorimeter response after one $\lam_T$ is
$E_{\rm calorimeter}=rE_{\rm ionization}+E_{\rm hadronic}$,
where $r\sim 1.6$ for an iron calorimeter.
Sixth, it is necessary to determine if the $\gl$-jet is charged
at appropriate points in the detector, and other analysis-dependent
criteria are satisfied, such that the $\gl$-jet
is identified as containing a muon. `Muonic' jets are discarded in the CDF
jets + missing energy analysis, but retained
in the corresponding OPAL analysis.
In the latter, the jet energy
of a jet that is `muonic' is computed as:
\beq
\ejet =p_{\rm jet}= 
E^{\rm tot}_{\rm calorimeter}+
\thetamuid(p_{\rm tracker}-\mbox{2 GeV})\,,
\label{muonjet}
\eeq
where $\thetamuid=1$ or 0, $p_{\rm tracker}=\mgl\beta\gam$
is the momentum as measured by the tracking system,
and the $2\gev$ subtraction is the energy
that would have been deposited by a muon in the calorimeter.

In the end, the
$\ejet=p_{\rm jet}$ as defined by the experiments
is normally quite different from the true gluino jet momentum,
and most events will be associated with large missing momentum.
Further, for moderately large $P$ (but not too close to 1), there
are large fluctuations on an event-by-even basis in how
the $\gl$-jets are treated.
Thus, one~\cite{bcg} employs
an event-by-event model of $\gl$ passage through
the detector accounting for $P$ at each hadronic collision and 
the associated calorimeter responses.

\begin{figure}
\begin{center}
{\psfig{figure=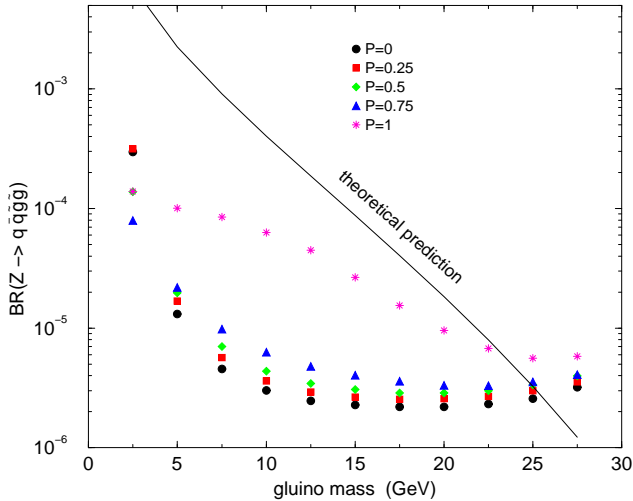,width=3.3in}}
\caption[]{We compare the prediction of $\br(Z\to q\anti q\gl\gl)$
compared to the extracted~\cite{bcg} OPAL 95\% CL upper limit
as a function of  $\mgl$ for $P=0,1/4,1/2,3/4,1$.
Both smearing and fragmentation effects are included.
Results are for the standard $\lam_T$ and $\vev{\Delta E}$
(SC1) case.~\cite{bcg}
\label{pscannormal}}
\end{center}
\end{figure}

\begin{figure}[h]
\begin{center}
{\psfig{figure=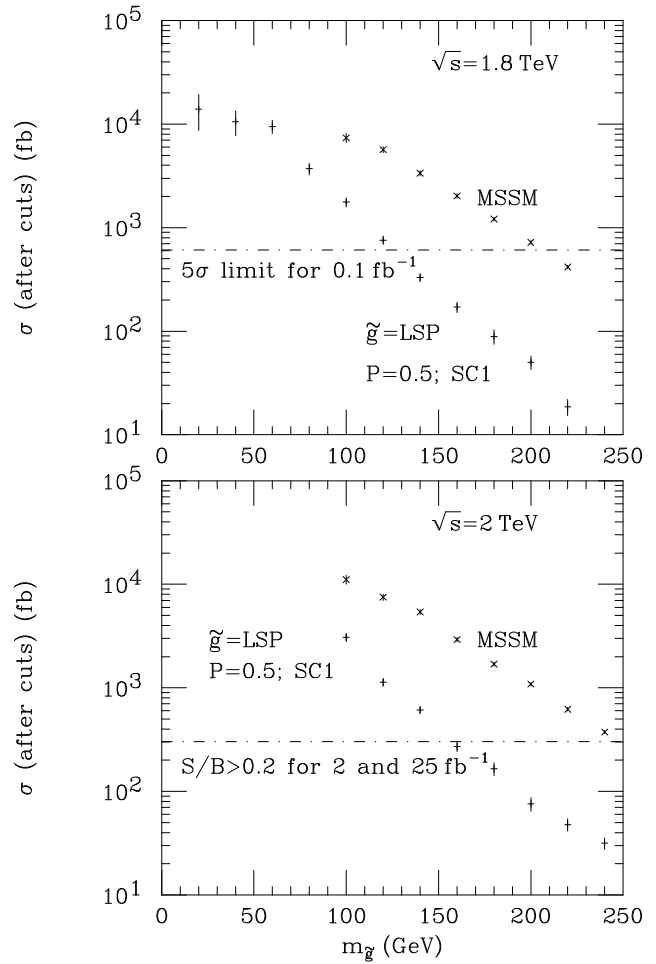,width=3.3in}}
\caption[]{The cross section (after cuts) in the ${\rm jets}+\ptmiss$ channel 
is compared to (a) the 95\% CL upper limit for $L=19\pbi$ (which
is the same as the $5\sigma$ signal level for $L=100\pbi$) at $\rts=1.8\tev$
and (b) the $S/B=0.2$ level at Run-II ($L\geq 2\fbi$, $\rts=2\tev$)
as a function of $\mgl$ for $P=1/2$.
Standard choices for $\lam_T$ and $\vev{\Delta E}$ are employed.
\label{phalfcdf}} 
\end{center}
\end{figure}

Sample results for OPAL and CDF are illustrated in Figs.~\ref{pscannormal}
and \ref{phalfcdf}. These figures illustrate that, for `standard'
choices~\cite{bcg} of $\lam_T$ and $\vev{\Delta E}$, 
one can use the jets + missing
energy OPAL and CDF analyses to exclude any
$\mgl$ from $\sim 3\gev$ up to $\sim 130-150\gev$, regardless of
the charged fragmentation probability $P$. For $P>1/2$, there is
some sensitivity to the $\lam_T$ and $\vev{\Delta E}$ scenario choices:
limits could be weaker (or stronger). For choices
that yield weak limits when
$P>1/2$, one can use the OPAL and CDF searches for tracks corresponding
to a heavily-ionizing charged particle to eliminate all $\mgl$ values
up to $\sim 130-150\gev$ except in the interval $23\lsim \mgl\lsim 50\gev$,
which is the gap between the OPAL analysis and the current version of
the CDF analysis. A refined CDF heavily-ionizing-track
analysis should be able to eliminate this gap.

\section{Delayed decay signals for gauge-mediated supersymmetry breaking 
(GMSB)}

The two canonical GMSB possibilities are:
$\staur$=NLSP, with $\staur\to\tau\gtino$; and $\cnone=$NLSP
followed by $\cnone\to\gam\gtino$, where the $\gtino$
is the Goldstino.  In either case, the NLSP decay
can be either prompt or delayed.  In the $\staur$-NLSP case, detection
of SUSY will be easy, either using heavily ionizing tracks \cite{fengmoroi}
for long path length of the $\staur$ or $\tau$ signals \cite{duttaetc}
if the decay is prompt. However, if the $\cnone$ is the NLSP, detection
of a SUSY signal can be much more challenging, and measurement
of the $\susyslash$ scale, $\sqrt F$, requires special 
attention.~\cite{chengun} In fact, it is quite possible, and required
in some models, that $\sqrt F\sim 1000-5000\tev$, in which case
\begin{equation}
(c\tau)_{\cnone=\wtil B\to \gam \gtino}\sim 130
\left({100\gev\over \mcnone}\right)^5
\left({\sqrt F\over 100\tev}\right)^4\mu {\rm m}
\label{ctauform}
\end{equation}
is typically quite large.
In particular, in GMSB models with a hidden sector communicating at two-loops
with a messenger sector, we have \cite{gunionbigf,murayamabigf}
(to within a factor of 5 or less)
$\sqrt F> \Lambda\sqrt f$,
where $f\sim 2.5\times10^4/g_m^4$ and $\Lambda$ is the parameter
that sets the scale of soft-susy-breaking masses. Roughly,
$40\tev\lsim\Lambda\lsim150\tev$ is required (see below), implying
$\sqrt F \gsim 1000-5000\tev$. Meanwhile, the gravitino has mass
$\mgtino={F\over \sqrt 3\mplanck}\sim 2.5
\left({\sqrt{F}\over 100\tev}\right)^2~{\rm eV}$, and
$\mgtino\lsim 1$ keV is preferred by cosmology, implying $\sqrt F\lsim
3000\tev$. Thus, we
should take seriously $1000\lsim\sqrt F\lsim 5000\tev$
and the possibility of delayed
$\cnone$ decays. At the very least, one should
explore the phenomenology of the model for the full
range of possible $\sqrt F$ values.

A recent study~\cite{chengun} has explored
Tevatron phenomenology for the full range of $\sqrt F$ in 
a sample model in which the superparticle masses have
the relative magnitudes typical of the simpler GMSB models with minimal
messenger sector content. In the model employed, the $\cnone$ is the LSP
and the sparticle masses are:
$\mcnone\sim 1.35\gev\cdot \Lambda$, $ \mcpone\sim
2.7\gev\cdot\Lambda$ ($\sim 2\mcnone$), $ \mgl\sim 8.1\gev\cdot\Lambda$
($\sim 6\mcnone$),
$\mslepr\sim 1.7\gev\cdot\Lambda$, $\mslepl\sim
3.5\gev\cdot\Lambda$ ($\sim 2\mslepr$), $\msq\sim11\gev\cdot\Lambda$ ($\sim
6\mslepr$), with $\Lambda$ in TeV. From these mass
formula, we see that if $\Lambda\lsim 40\tev$ then the $\slepr$
would have been seen at LEP or LEP2, while if $\Lambda\gsim 150\tev$
the $\gl$ and the $\sq$'s becomes so heavy that naturalness
problems for the Higgs sector would certainly be substantial.
For the above hierarchy of masses, 
the primary normal SUSY signal at the Tevatron
is the tri-lepton signal.
It is found~\cite{chengun} that this
signal is viable for $\Lambda\lsim 65\tev$
for any $\sqrt F$; but it does not distinguish
a SUGRA-like model from a GMSB model.
In order to distinguish between the two model possibilities, one
must detect the photon(s) that result from the $\cnone$ decays.

\begin{figure}
\psfig{figure=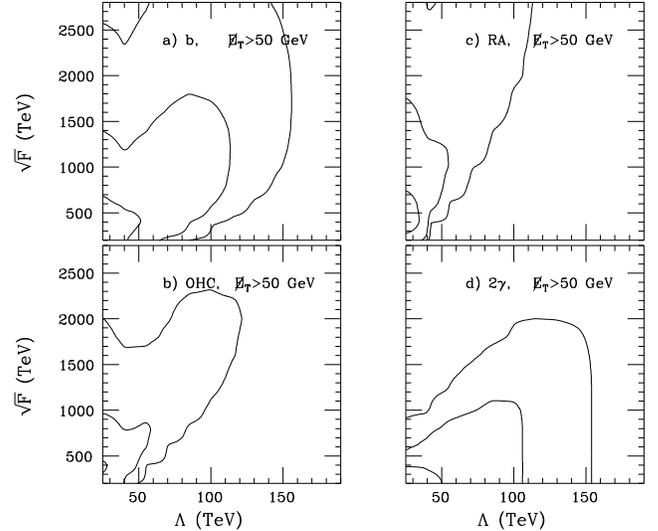,width=3.3in}
\caption{
$\sigma$ contours in the $(\protect\sqrt F,\Lambda)$
parameter space for the (a) impact-parameter ($b$), (b) outer-hadronic (OHC),
(c) roof-array (RA), and (d) prompt-two-photon ($2\gam$) signals.
Contours are given at $\sigma=0.16$, 2.5, and $50\fb$ --- these
correspond to 5 events at $L=30,2,0.1\fbi$ as for
Tev33, Run-II and Run-I, respectively.
\label{contoursfullfigv}}
\end{figure}

One possibility is to detect a 
prompt photon in association with the tri-lepton signal.
One finds that this will be possible only if
$\sqrt F$ is not very large.
Additional associated-photon signals that can be considered 
include: observation of a photon with
non-zero impact-parameter ($b$); decay of the $\cnone$
leading to an isolated energy deposit in an outer-hadronic-calorimeter cell
(OHC); a photon signal in a specially designed
roof-array detector placed on the roof
of the detector building (RA); and the appearance of two prompt (emergence 
before the electromagnetic calorimeter) photons
($2\gam$). The first three are present only if the $\cnone\to \gam\gtino$
decay is delayed, while the latter signal will be very weak if
the decay is substantially delayed.  After imposing strong
cuts that hopefully reduce backgrounds to a negligible level (detailed
detector studies being needed to confirm), the regions
in $(\sqrt F,\Lambda)$ parameter space for which these signals
are viable at the D0 detector 
for Run-I, Run-II and Tev33 luminosities at the Tevatron
are illustrated in Fig.~\ref{contoursfullfigv}.

We can summarize as follows.  If both $\sqrt F$ and $\Lambda$ are large,
then we will not see either the tri-lepton signal or the prompt $2\gam$
signal. However, Fig.~\ref{contoursfullfigv} shows that 
the large impact parameter photon signal
from delayed decays of the $\cnone$ can cover most
of the preferred parameter space region not accessible by the former
two modes.
The roof-array detector also provides an excellent signal at large $\sqrt F$.
Putting all the signals together, the portion of $(\sqrt F,\Lambda)$ 
parameter space for which
a GMSB $\cnone=$NLSP SUSY signal can be seen at the Tevatron
is greatly expanded by delayed $\cnone$ decay signals.
We re-emphasize the fact that one needs
the $b$, RA and/or OHC delayed-decay signals to distinguish
GMSB from a SUGRA model with GMSB-like boundary conditions
when $\sqrt F$ is too large for a viable prompt photon signal.
Finally, if delayed-decay signatures are seen,
an absolutely essential goal will be to determine $\sqrt F$
(the most fundamental SUSY parameter of all): the $b$ and RA signals
provide the information needed to do so.

\section{The nightmare R-parity violating (RPV) scenario}

This scenario~\cite{jfgnm} is designed as a warning against complacency
regarding SUSY discovery.
The  first ingredient in the nightmare scenario is a non-zero B-violating RPV
coupling (often denoted $\lam^{\prime\prime}$), 
which leads to LSP decay to three jets: $\cnone\to 3j$.
This means that the $\cnone$'s, resulting
from chain decay of a pair of produced supersymmetric particles, do not yield
missing energy. The standard jets + $\ptmiss$ signal is absent.
Of course, for universal gaugino masses at the GUT scale
this is not a problem since $\mcpmone\sim 2\mcnone$
(at energy scales $< 1\tev$). For such large $\mcpmone-\mcnone$,
a robust signal for supersymmetry
is provided~\cite{bingun}
by the like-sign dilepton signal that arises when two
$\cpone$'s (or two $\cmone$'s) are produced in the decay
chains and both decay leptonically: \eg\ $\cpone\cpone\to
\ell^+\ell^+\nu_\ell\nu_\ell \cnone\cnone$. Since the leptons have significant
momentum and the neutrinos yield some missing momentum,
the like-sign lepton events are typically quite easily
isolated at the LHC, and for lower SUSY mass scales, also at the
Tevatron.~\cite{bingun,likesignrviol}

However, as discussed in an earlier section,
if gaugino masses are not universal it is very possible
to have $M_2<M_1\ll|\mu|$ (at energy scales $<1\tev$), 
leading to both $\cpmone$ and $\cnone$
being SU(2) winos  with $\mcpmone\simeq\mcnone$. Also,
in the models with $|\mu|\ll M_{1,2}$
the lightest {\it two} neutralinos and the lightest chargino
are {\it all} closely degenerate: $\mcnone\simeq\mcpmone\simeq\mcntwo$.
In either case, the
leptons in $\cpmone\to\ell^{\pm}\nu\cnone$ decays are very soft.
The like-sign dilepton
signal would be very weak (after necessary cuts requiring
reasonable momenta for the leptons). The implications
of these scenarios are the following.
At LEP2 one would need to use the $\epem\to\gam\cpone\cmone$
photon-tag signal, but, unlike the case where the $\cnone$'s
from $\cpmone$ decays
yield missing energy, the $\cpone\cmone$ would
decay to a final state containing six (relatively soft) jets.
Perhaps, such a signal can be shown to be viable over backgrounds
up to some reasonable value of $\mcpmone$. To go beyond this
value would require a viable signal at the Tevatron and/or LHC.
However, at the hadron colliders, leptonic signals will
be very weak. Aside from $W$ decays, energetic leptons 
can emerge only from decays of the heavy gauginos
(\eg\ the higgsino states in the $M_2<M_1\ll|\mu|$ case) 
that are present by virtue of either being directly produced
or arising in decays of still heavier produced supersymmetric
particles. If the leptonic signals turn out to
be too weak, the only signal with a substantial rate
will be spherical events containing an extra large number
of jets. This signal might prove very difficult to isolate from
backgrounds.

\section{Two topics regarding Higgs bosons in SUSY.}

The first topic concerns the use of
experimental limits on the lightest SUSY Higgs boson, $\hl$,
to exclude parameter regions in various SUSY models.
The second topic is the construction of a truly difficult
scenario for SUSY (or any) Higgs detection.

\subsection{Model constraints from limits on the $\hl$}

To illustrate the possibilities, I present a brief discussion
of two representative papers.  The first is that of de Boer
\etal.~\cite{deboer} They assume universal mSUGRA CMSSM (constrained
MSSM) boundary conditions
and impose radiative electroweak symmetry breaking and gauge
coupling unification with $\alpha_s(\mz)=0.122$
and $\mt=173.9\pm 5.2\gev$. They also require
$b-\tau$ Yukawa unification, with $m_b(m_b)=4.2\pm0.15\gev$ 
(do we really know it so well?). Additional input data is
the current combined ALEPH/CLEO result for $b\to s\gam$ (including
combined errors) and Higgs mass limits from LEP2. Regarding
the latter, the CMSSM approach with RGE electroweak symmetry
breaking implies that $\mha$ is large and that the $\hl$ is
very SM-like. Thus, they require $\mhl\gsim 89\gev$ at 95\% CL.
Finally, they require $\Omega h^2<1$ for the relic neutralinos
of the model. With this input, including a systematic treatment
of experimental errors, they compute the $\chi^2$ for
different parameter choices in the CMSSM context.

They find significant constraints on the allowed parameter space.
It is convenient to think of the allowed parameter regions as follows.
First, by imposing $b-\tau$ unification they end up with only
4 good $\tanb$ and sign($\mu$) solution scenarios: two
at low $\tanb$ and two at high $\tanb$.  These 4 possibilities
are then restricted by other constraints as shown in the following Table.

\begin{table}[h]
\normalsize
\begin{center}
\begin{tabular}{|c|c|c|}
\hline
Constraint & $\tanb=1.65$ & $\tanb=1.65$ \cr
\ & $\mu<0$ & $\mu>0$ \cr
\hline
$b\to s\gam$ & OK & OK \cr
$\mhl>90$? & No & $\mhalf>400$ \cr
$\Omega h^2<1$? & $m_0<300$ & $m_0<300$ \cr
\hline
Constraint  &  $\tanb=35$ & $\tanb=64$ \cr
\ & $\mu<0$ & $\mu>0$ \cr
\hline
$b\to s\gam$ & $\mhalf>700$ & $\mhalf>500$ \cr
$\mhl>90$? & Yes & Yes \cr
$\Omega h^2<1$? & $m_0>300$ & $m_0>300$ \cr
\hline
\end{tabular}
\end{center}
\end{table}

The Higgs limits are most restrictive for the low-$\tanb$ solutions.
First, for low $\tanb$, $\mu<0$ is pretty much excluded unless one allows
$m_0,\mhalf>1\tev$. Second,
low $\tanb$ and $\mu>0$ will soon be excluded if no Higgs is seen at
LEP200: for $\mu>0$ and $m_0,\mhalf<1\tev$, the other constraints imply
$\mhl^{\rm max}=97\pm6\gev$ (error dominated by uncertainty in $\mt$).
If $\tanb$ is large, then
$\mhl^{\rm max}=120\pm 2\gev$, which will hopefully be testable at TeV33.
The best $\chi^2$ solutions have large squark masses $>1\tev$ and fine-tuning
problems.

Many other studies, especially in the fixed-point context, 
reach very similar conclusions. For example, Carena \etal~\cite{carena}
show that $\mhl>89\gev$ implies a lower bound on $\tanb$ well above 
the perturbativity bound unless the stop mass matrix is carefully chosen.
In particular, the fixed-point value of $\tanb\sim 1.5$ is
allowed only if the 
heavier stop is not too heavy (\ie\ there is an implicit upper bound
on $\mstoptwo$).
If the bound on $\mhl$ increases to $\mhl\gsim 103\gev$ (as expected
at LEP200), the low-$\tanb$ fixed point scenario will be ruled out.
Of course, one should keep in mind
that the low-$\tanb$ fixed point solution is ruled out in mSUGRA and CMSSM
if we require $\Omega h^2<1$~\cite{ellisetal} and/or no charge/color
breaking.~\cite{abeletal} This means, that we can only have
a low-$\tanb$ fixed-point solution with $\mhl>89\gev$ that is
consistent with $\Omega h^2<1$ if we disconnect the slepton, Higgs
and squark soft-supersymmetry-breaking mass parameters
by not requiring a universal value at the GUT scale.
Finally, we recall that adequate electroweak baryogenesis
in the MSSM requires $\mhl$ in the LEP2 range, 
a light $\stopone$, and small stop mixing. Imposing these
constraints in conjunction with the fixed-point low-$\tanb$ solution
requires $\mstoptwo>2\tev$.

Of course, all these constraints
depend greatly upon the fact that the MSSM contains exactly
two doublets and no other Higgs representations.  For example,
the constraints found in these studies 
are obviated if one adds extra singlet(s)
$S$ with $\lam SH_1H_2$ style coupling.

\subsection{A very difficult Higgs scenario: is there a no-lose theorem?}

An interesting question that has emerged in several different
models is the question of whether there is a no-lose theorem
for Higgs discovery at a $\rts=500\gev$ $\epem$ collider.
Typically,~\cite{kimoh,espgun} 
if one adds just one or two Higgs bosons to the
spectrum the answer is yes: one or more of the scalar Higgs
bosons will be discovered in the $Z\h$ production mode.
However, the situation could be much more complicated.
A very difficult case~\cite{espgun} is one in which there are
many Higgs bosons, as could arise in a string
model with many U(1)'s,~\cite{dienes}
and they share the $ZZ$-Higgs coupling-strength-squared ($g_{ZZ\h}^2$)
fairly uniformly. Further, assume these Higgs bosons are spread out
such that the experimental resolution is insufficient to resolve
the separate peaks, in which case
the only signal is an unresolved continuum excess over background.
Finally, assume the Higgs bosons all decay into a variety of channels,
including invisible decays, various $q\anti q$ channels, \etc,
in which case identification of the $\h$ decay final state would 
not be useful because of the large background in any one channel.
In particular, in $\epem\to Zh$, there would be
no guarantee we can use $Z\to q\anti q$ or $\nu\anti\nu$ decays 
because of the large number of possible channels in the recoil state
and, thus, small signal relative to background in any one channel.
The only clearly reliable signal would be an excess in the recoil
$\mx$ distribution in $\epem\to Z X$ (with $Z\to\epem$ and $\mupmum$).

To describe this scenario quantitatively, one~\cite{espgun}
employs a continuum description in which there
are Higgs bosons from $\mhmin$ to $\mhmax$.
Defining $K(\mh)$ as the $g_{ZZh}^2$ strength (relative to SM strength)
as a function of $\mh$,
one then can write two sum rules:
\bea
&\int& d\mh K(\mh)\geq 1 \label{ksumint}\\
&\int& d\mh K(\mh)\mh^2\leq \vev{M^2}\,,
\label{msqlimint}
\eea
where the former becomes an equality if only Higgs singlet and doublet
representations are involved.
The key to a no-lose theorem is to limit $\vev{M^2}$.
In the context of supersymmetry one can write
$\vev{M^2}\equiv m_B^2=\lam v^2$, where $v=246\gev$
and $\lam$, a typical quartic Higgs coupling at low energy scales, 
is limited by requiring perturbativity for $\lam$ up to some high scale
$\Lambda$. In the most general SUSY model, one
finds $m_B\leq 200\gev$ for $\Lambda\sim 10^{17}\gev$.
Alternatively, independently of a SUSY context,
the success of fits to precision electroweak data
using $\mhsm\lsim 200\gev$ implies,
in a multi-Higgs model, that the Higgs bosons with large $K(\mh)$
must have average mass $\lsim 200\gev$, which would imply $m_B\lsim 200\gev$.

Taking $K(\mh)=K$, a constant, Eq.~(\ref{ksumint}) 
leads to $K=1/(\mhmax-\mhmin)$ (assuming only singlet and doublet
representations), and Eq.~(\ref{msqlimint}) implies
\bea
m_B^2\equiv\vev{M^2}&\geq&
{1\over 3}\left([\mhmax]^2+\mhmax\mhmin+[\mhmin]^2\right)\,.
\label{newlim}
\eea
The maximal spread is achieved for $\mhmin=0$, in which case
Eq.~(\ref{newlim}) requires $\mhmax\leq \sqrt 3 m_B\leq 340\gev$.  

To analyze this situation,~\cite{espgun} assume $\rts=500\gev$, 
for which $\sigma(Z\h)$ for a SM-like $\h$ is substantial out to 
$\mh\sim 200\gev$. [$\sigma(Z\hsm)$ falls from $70\fb$ at low $\mh$ to $42\fb$
at $\mh\sim 200\gev$.] Confining the signal region to $70\leq\mh\leq 200\gev$,
a fraction $f\sim 0.4$ of the uniform $K(\mh)$ spectrum would lie
in this region. If LEP2 data can eventually be used to show that $K(\mh)$
is small for $\mh\lsim 70\gev$ (\ie\ $\mhmin=70\gev$) then
$\mhmax=300\gev$ [from Eq.~(\ref{newlim}) with $m_B=200\gev$]
and a fraction $f\sim 0.55$ would lie in the $70-200\gev$ region.
Alternatively, one can consider only the $100-200\gev$ interval (to
avoid the large background in the vicinity of $\mh\sim\mz$), in which case
$f\sim 0.3$ for $\mhmin=0$ and $f\sim 0.43$ for $\mhmin=70\gev$, respectively.

The results for the overall excess in $Z\h$, with $Z\to\epem+\mupmum$,
integrated over the $70-200\gev$ and $100-200\gev$ intervals, assuming
$\int K(\mh)=1$ over the interval, are given in Table~\ref{ghsrates},
assuming an integrated luminosity of $L=500\fbi$ (which is very optimistic).
Including the factor $f$, one finds $S\sim 1350 f$ with a background
of either $B=6340$ or $B=2700$, for the $70-200\gev$ or $100-200\gev$ windows,
respectively. Correspondingly, one must
detect the presence of a broad $\sim 21\%f$ or
$\sim 50\%f$ excess over background, respectively.
For $f\sim 0.4 - 0.55$ in the 1st case and $f\sim 0.3-0.43$
in the 2nd case, this would probably be possible.
Nominally, $S/\sqrt B\sim 17f$ and $\sim 26f$ for the $70-200\gev$
and $100-200\gev$ windows in $\mx$, respectively.
However, if $L\lsim 200\fb$, the detection of the excess will
become quite marginal.
As an aside, we note~\cite{espgun} that $\epem\to\epem \h$ via
$ZZ$-fusion is not useful because of very small $S/B$.

\begin{table}[h]
\caption[fake]{Approximate $S$, $B$ and 
$S/\sqrt B$ values for $Z\h$ (with $Z\to\epem+\mupmum$)
after integrating the $\mx$ recoil mass spectrum 
from (a) 70 GeV to 200 GeV and (b) 100 GeV to 200 GeV, assuming
that many Higgs bosons are distributed evenly
throughout the interval with uniform $K(\mh)$. 
Results are for $\rts=500\gev$, $L=500\fbi$.}
\medskip
\begin{center}
\begin{tabular}{|c|ccc|ccc|}
\hline
 $\mx$  & \multicolumn{3}{c|} {$Z\h$, $Z\to\epem+\mupmum$} \\
 Interval  & $S$ & $B$ & ${S/\sqrt B}$ \\
\hline
$70-200$ & 1350 & 6340 & 17 \\
$100-200$ & 1356 & 2700 & 26 \\
\hline
\end{tabular}
\end{center}
\label{ghsrates}
\end{table}

\begin{table}[h]
\caption[fake]{Approximate $S$, $B$ and 
$S/\sqrt B$ values for $Z\h$ (with $Z\to\epem+\mupmum$) 
in each of the thirteen 10 GeV bins in $\mx$
from 70 to 200 GeV, assuming that $S\sim 1350$ events
are distributed equally among these bins.
We assume $\rts=500\gev$, $L=500\fbi$.}
\medskip
\begin{center}
\begin{tabular}{|c|c|c|c|c|}
\hline
Bin No. & 1 & 2 & 3 & 4 \\
$S$ & 104 & 104 & 104 & 104 \\
$B$ & 1020 & 1560 & 1440 & 734 \\
$S/\sqrt B$ & 3.3 & 2.6 & 2.7 & 3.8 \\
\hline
Bin No. & 5 & 6 & 7 & 8--13 \\
$S$ & 104 & 104 & 104 & 104 \\
$B$ & 296 & 162 & 125 & $\sim 130$ \\
$S/\sqrt B$ & 6.0 & 8.2 & 9.3 & 9.1 \\
\hline
\end{tabular}
\end{center}
\label{ghsbin}
\end{table}

Of course, if an excess is observed, the next interesting question
is whether we can analyze the amount of this excess on a bin-by-bin basis.
The situation is illustrated in Table~\ref{ghsbin}
assuming that the roughly 1350 (\ie\ $f=1$ for the moment)
signal events are distributed
equally in the thirteen 10 GeV bins from 70 to 200 GeV.
Table~\ref{ghsbin} gives $S$ for $f=1$, 
$B$ and the corresponding $S/\sqrt B$ value for each bin.
Both $S$ and $S/\sqrt B$ must be reduced by $f$.
One sees that $L=500\fbi$ would 
yield $S/\sqrt B > 3$ only for the $\mx\gsim 120\gev$ bins
when $f\sim 0.5$. 
Further, with only $L=100\fbi$ (as might be
achieved after a few years of running at a `standard' luminosity design),
this bin-by-bin type of analysis would not be possible for 
10 GeV bins if $f\sim 0.5$; one really needs $L=500\fbi$.

A final question is how many Higgs force us into the continuum scenario?
In the inclusive $\epem\to ZX$ mode, with $Z\to \epem,\mupmum$,
the electromagnetic calorimeter and tracking resolutions
planned for electrons and muons imply $\Delta m\sim 20\gev$ at $\rts=500\gev$.
As a result, something like five Higgs bosons distributed from 70 to 200 GeV
would put us into the continuum scenario unless a specific Higgs decay final
state (for which resolutions are expected to be below $10\gev$
and backgrounds would be smaller)
could be shown to be dominant.

\section{Doubly-charged Higgs and higgsinos in 
supersymmetric L-R models}

In supersymmetric L-R symmetric models, the Lagrangian
cannot contain terms that explicitly violate R-parity. The
presence or absence of RPV is determined by whether or not
there is spontaneous RPV. There are two generic
possibilities.~\cite{mohlrtheory}

If certain higher dimensional operators are small or absent, then
the scalar field potential must be
such that L-R symmetry breaking induces RPV
through some combination of non-zero $\vev{\wtil\nu^c_i}$'s. In this
case, the $W_R$ mass scale is low and, of course, there are
lots of new phenomena associated with RPV. 
In this scenario, the triplet Higgs and higgsinos, 
including $\dmm_{L,R}$ and their fermionic partners, are not necessarily light.
Considerable phenomenological discussion of the resulting RPV
signatures for this case has appeared.~\cite{huitu}

If the above-mentioned higher-dimensional operators are present and
are of full strength (but, of course, $\propto 1/\mgut$ or $1/\mpl$),
then L-R symmetry breaking does not require RPV. 
In this case, the $W_R$ mass scale must be very large. 
Further, the $\Delta_L$ triplet members and their superpartners
must be very heavy unless one removes the
(naturally present) parity-odd singlet from the theory
(which is normally included in order to avoid
$v_L\neq 0$ vacua). 
However, when the R-sector Higgs mechanism comes in at high
scale (assumed to be above the SUSY breaking scale) 
to give $v_R\neq0$ and generate $W_R$ mass, one is breaking
a U(3) symmetry and there are 4 surviving massless (goldstone) fields,
which are the $\dmm$ superfield and its charge conjugate, whose component
fields only become massive via the higher-dimensional operators.
In this case, it is natural for
the mass scales of the $\dmm_R$ and $\wtil\dmm_R$, $\sim v_R^2/\mpl$,
to be at the $\sim 100\gev$ level. 

The phenomenology of doubly charged Higgs bosons
has a long history.~\cite{jfgdmm} The above $\dmm_R$ (hereafter
we drop the $R$ subscript) would generally be
narrow. Noting that $\dmm\to \wm\dm$ 
is expected to be kinematically forbidden,
its primary decay modes would most probably 
be via the Majorana couplings associated
with the see-saw mechanism for neutrino mass generation:
\begin{equation}
{\cal L}_Y=ih_{ij}\psi^T_{i} C\tau_2\Delta\psi_{j}+{\rm h.c.}
\,,
\label{couplingdef}
\end{equation}
where $i,j=e,\mu,\tau$ are generation indices, and
$\Delta$ is the $2\times 2$ matrix of Higgs fields:
\begin{equation}
\Delta=\pmatrix{\delp/\sqrt{2} & \dpp \cr \hzero & -\delp/\sqrt{2} \cr}\,.
\end{equation}
Limits on the $h_{ij}$ by virtue of the $\dmm\to
\ell^-\ell^-$ couplings include: Bhabbha scattering, $(g-2)_\mu$, 
muonium-antimuonium conversion, and $\mu^-\to e^- e^- e^+$.  
Adopting the  convention
\begin{equation}
|\hdmm_{\ell\ell}|^2\equiv c_{\ell\ell} \mdmm^2(\gev)\,,
\label{hlimitform}
\end{equation}
one finds $c_{ee}< 10^{-5}$ (Bhabbha) and $\sqrt{c_{ee}c_{\mu\mu}}<10^{-7}$
(muonium-antimuonium) are 
the strongest of the limits. There are no limits on $c_{\tau\tau}$
which is, naively, expected to be the largest.
If all the $c$'s are very tiny, virtual versions of $\dmm\to\dm\wm$ 
could be important.

Regarding production, because of the very large $W_R$ mass, 
the doubly-charged Higgs bosons would be primarily
produced at hadron colliders via $\gam^*,Z^*\to\dmm\dpp$.
At an $\emem$ or $\mu^-\mu^-$ collider they could be produced
directly as an $s$-channel resonance via
the lepton-number-violating couplings $h_{ee}$ and $h_{\mu\mu}$, respectively.
The strategy for discovering and studying the $\dmm$ would be
the following.
First, one would discover the $\dmm$ in
$p\anti p\to \dmm\dpp$ with $\dmm \to
\ell^-\ell^-,\dpp\to\ell^+\ell^+$ ($\ell=e,\mu,\tau$) at TeV33 or
LHC.~\cite{glp} One finds that
$\dmm$ detection at the Tevatron
($\rts=2\tev$, $L=30\fbi$) is possible for $\mdmm$ 
up to $300\gev$ for  $\ell=e$ or $\mu$ and  up to $180\gev$ for $\ell=\tau$.
At the LHC, $\dmm$ discovery is possible up to roughly $925\gev$ ($1.1\tev$)
for $\ell=e,\mu$ and $475\gev$ ($600\gev$) for $\ell=\tau$,
for $L=100\fbi$ ($L=300\fbi$).    
Thus, TeV33 + LHC will tell us if such a $\dmm$ exists in the mass
range accessible to the next linear collider or a first muon collider,
and, quite possibly, its decays will
indicate if it has significant coupling to $e^-e^-$
and/or $\mu^-\mu^-$ (unless $\tau^-\tau^-$ is completely dominant,
as is possible). Whether or not these decays are seen,
we will wish to determine the strength of these couplings
by studying $e^-e^-$ and $\mu^-\mu^-$ $s$-channel production
of the $\dmm$. We note that
if the $\dmm$ is observed at the LHC, we will know ahead
of time what final state to look in and have a fairly good
determination of $\mdmm$.

At the NLC, taking $L=50\fbi$ and defining $R$ to be the beam energy
spread in percent,
\begin{equation}
N(\dmm)\sim 3\times 10^{10}\left({c_{ee}\over 10^{-5}}\right)
\left({0.2\%\over R}\right)\,,
\label{eventratenarrow}
\end{equation}     
implying an enormous event rate if $c_{ee}$ is near its upper bound.
The ultimate sensitivity to $c_{ee}$ when $\gamdmm$ is much
smaller than the beam energy spread 
can be estimated by supposing that 100 events are required. From
Eq.~(\ref{eventratenarrow}), we predict 100 $\dmm$ events for
\begin{equation}
\left.c_{ee}\right|_{\rm 100~events}\sim 3.3\times 10^{-14} (R/0.2\%)
\,,
\end{equation} 
independent of $\mdmm$, which is dramatic sensitivity.  
Because of the much smaller $R$ values possible at a $\mu^-\mu^-$ 
collider ($R\sim 0.003\%$ is possible), comparable or greater sensitivity
to $c_{\mu\mu}$ could be achieved there despite the lower expected
integrated luminosity.

In the L-R symmetric models the phenomenology of the
doubly-charged Higgsinos would be equally interesting.~\cite{mohlr}
The basic experimental signatures always involve $\tau$'s.
In non-GMSB SUSY,
if $h_{\tau\tau}$ is full strength ($\sim 0.5$) then it influences
the RGE's so that the $\stau$'s (especially $\staur$) 
are lighter than $\wtil e$ and $\wtil \mu$, even if $\tanb$ is not large.
Further, starting with a common mass at
the $v_R$ scale, evolution leads to $m_{\dmm}<m_{\wtil\dmm}$
and the $\dmm$ would be easily visible as described above.
Less attention has been paid to $\wtil\dmm,\wtil\dpp$, which could be produced
at the Tevatron in pairs. Indeed, for $\mdmm=m_{\wtil\dmm}$,
the $\wtil\dmm\wtil\dpp$ pair cross section is bigger than
that for $\dmm\dpp$ due to the fact that the former
is not $p$-wave suppressed. Normally,
$\wtil\dmm\to \staur\tau$ is kinematically allowed
and will dominate over all other lepton channels because of larger coupling.
The dominant $\staur$ decay would be $\staur\to\cnone\tau$.
Thus, a typical signature would be
$p\anti p\to\wtil\dmm\wtil\dpp\to
\tau^-\tau^-\tau^+\tau^+ +\ptmiss$.
Note that the presence of $\ptmiss$ would make
reconstruction of the $\dmm$ and $\dpp$ masses difficult.

In the GMSB context there are some alterations to the above
scenario. First, one finds that
the $\wtil\dmm$ is now lighter than the $\dmm$. 
In fact, the $\wtil\dmm$ could even be the NLSP.
If not, the $\staur$ very probably is (even for minimal messenger
sector content), with $\staur\to \tau\gtino$ being its dominant decay.
The typical signature would be the same as above except 
the $\ptmiss$ would now be due to the $\gtino$'s rather than $\cnone$'s.
In the small portion of parameter space where the $\cnone$ is the NLSP,
the signature for $\wtil\dmm\wtil\dpp$ production changes to 
$4\tau 2\gam+\ptmiss$, where the $\gam$'s come from the $\cnone\to\gam\gtino$
decays.

Overall, a supersymmetric L-R symmetric model would give rise
to a very unique phenomenology with many exciting ways to explore
the content and parameters of the model.

\section{Conclusion}

I have tried to give an overview of recent results
in supersymmetry phenomenology with emphasis on unusual scenarios
that one might encounter, especially ones for which detection
of supersymmetric particles and/or the SUSY Higgs bosons might 
require special experimental/analysis techniques.  Experimentalists
should pay attention to these special cases to make sure that
their detector designs, triggering algorithms and analysis techniques
do not discard these possibly important signals.

\section*{Acknowledgements}
This work was supported in part by the U.S. Department of Energy.

\end{document}